\documentclass[aps,aps,twocolumn]{revtex4}
\usepackage{graphicx}
\usepackage{dcolumn}
\usepackage{bm}
\usepackage{amsmath}
\usepackage{txfonts}
\usepackage[scaled]{helvet}
\usepackage{color}
\usepackage[breaklinks=true,colorlinks,citecolor=blue,linkcolor=blue,urlcolor=blue]{hyperref}
\usepackage{makecell}
\usepackage{booktabs}
\usepackage{threeparttable}

\begin{document}

\title{Rashba spin splitting based on trilayer graphene systems}
\author{Xinjuan Cheng, Liangyao Xiao and Xuechao Zhai$^*$}
\affiliation{Department of Applied Physics and MIIT Key Laboratory of Semiconductor Microstructures and Quantum Sensing, Nanjing University of Science and Technology, Nanjing 210094, China}
\email{zhaixuechao@njust.edu.cn}
\date{\today}

\begin{abstract}
We establish a general Rashba Hamiltonian for trilayer graphene (TLG) by introducing an extrinsic layer-dependent Rashba spin-orbit coupling (SOC) arising from the off-plane inversion symmetry breaking. Our results indicate that the band spin splitting depends strongly on the layer-distribution and sign of Rashba SOC as well as the ABA or ABC stacking order of TLG. We find that spin splitting is significantly enhanced as the number of layers of the Rashba SOC with the same sign and magnitude increases. For the spatially-separated two Rashba SOCs of the same magnitude but the opposite sign, no spin splitting arises in ABC-TLG due to the preservation of inversion symmetry that ensures the complete cancellation of contributions from the opposite layers, whereas nonzero spin splitting is observed for ABA-TLG due to its own lack of inversion symmetry. We further illustrate that gate voltage is effective to modulate the spin-polarized states near the band edges. Moreover, we use density functional theory calculations to verify the Rashba splitting effect in the example of TLG interfaced by Au layer(s), which induce simultaneously the effective terms of Rashba SOC and gate voltage. Our results demonstrate the significance of layer and symmetry in manipulating spin and can be extended to multilayer graphene or other van der Waals interface systems.
\end{abstract}

\maketitle

\section{Introduction}

Flat graphene in its natural state has no magnetism and negligible spin-orbit couplings (SOCs) \cite{CasNet,KonGmi,SiPra}, resulting in its very limited application in spintronics despite its excellent mechanical, thermal and electronic properties \cite{CasNet,NovFal}. Because the spin degeneracy of electrons in graphene inhibits its development in spintronics, driving spin splitting has long been a key goal for scientists and engineers \cite{Yazyev,PesMacs,AvsTan,HanKaw,DayRay,ZolGmi2020,AvsOch}. In experiment, it is effective to open the spin degeneracy by inducing magnetism or SOCs in graphene with edge engineering \cite{Magda,Slota}, adatoms \cite{GonHer} or proximity effect \cite{GhiKav,AvsOch}. For SOCs, it usually includes \cite{SieFab,KanMe} Rashba type, Ising type (valley-Zeemann term) and Kane-Mele type {\it etc}, among which Rashba SOC is easier to induce through interface engineering since only the off-plane ($z\rightarrow-z$) inversion symmetry \cite{ManKoo} needs to be broken.

In recent years, numerous studies have been focused on the Rashba physics based on graphene monolayer, bilayer and multilayer \cite{DedFon,Rashba,MirSch,MarVar2012,Zhai2014,WangKi,ManKoo,MarVar2016,FarTan,KriGol,
YangLoh,SinEsp,LopCol,PerMed,GhiKav,SieFab,zhai2022,QiaoLi,Hoque,Khokh,TiwSri,LiZhang} (see Table 1 for the experimentally-reported Rashba interface systems). The key electronic property induced by Rashba SOC is spin chirality in band (spin vortex in momentum space) \cite{Rashba,Zhai2014}, which is an intriguing form of spin splitting. Remarkably, a large Rashba splitting up to 0.1~eV was experimentally achieved by contacting graphene with Au layer \cite{MarVar2012}, unlike graphene/Ni where magnetism is induced besides Rashba SOC \cite{DedFon,PerMed}. Inspiringly, room-temperature spin Hall effect induced by Rashba spin splitting has been observed in graphene-based heterostructures \cite{GhiKav2019,BenTor}. Nevertheless, we have very little knowledge about the Rashba effect in trilayer graphene (TLG) till now, especially considering its stacking order, ABC type in Fig.~1(a) or ABA type in Fig.~1(b), between which significant differences exist in electrically-controllable electronic properties \cite{LuiLi,BaoJing,JhaCra,KhoKhr}, spin proximity effect \cite{ZolGmi2022} and second harmonics \cite{ShaLi}.

\begin{table} \label{T1}\footnotesize
\centering
\caption{Interface engineered Rashba SOC (strength $\lambda$) in graphene-based hybrid systems available in the existing references. Here, we are mainly concerned with the cases where Rashba SOC plays a dominant role in the spin-dependent interactions. The abbreviations MLG, BLG and FLG denote monolayer, bilayer, and five-layer graphene, respectively.}

\begin{threeparttable}
\begin{tabular}{lllll}
\toprule[1pt]
   \quad Interface systems \quad &\quad\quad{Strength ($\lambda$)} &\quad\quad{Experiment}&\quad\quad{Theory}\\

   \hline
\quad MLG/Ni(111)&\quad\quad{$\sim10^2$~meV}&\quad\quad {\cite{DedFon}}&\quad\quad {\cite{PerMed}}\\
\quad MLG/Au &\quad\quad{7-10$^2$~meV} &\quad\quad{\cite{MarVar2016,FarTan,MarVar2012}}&\quad\quad {\cite{KriGol,LopCol}}\\
\quad MLG/MoTe$_2$&\quad\quad{2.5~meV}&\quad\quad {\cite{Hoque}}&\quad\quad {\cite{Hoque}}\\
\quad MLG/Bi$_{0.3}$Sb$_{1.7}$Te$_3$&\quad\quad{$\sim$10~meV}&\quad\quad {\cite{Khokh}}\\
\hline
\quad BLG/WSe$_2$ (biased) &\quad\quad{10-10$^2$~meV} &\quad\quad {\cite{TiwSri}}\\
\hline
\quad FLG/2H-TaS$_2$ &\quad\quad{70~meV} &\quad\quad {\cite{LiZhang}}&\quad\quad {\cite{LiZhang}}\\
\bottomrule[1.0pt]
\end{tabular}
      \end{threeparttable}
\end{table}

\begin{figure}
\centerline{\includegraphics[width=8.5cm]{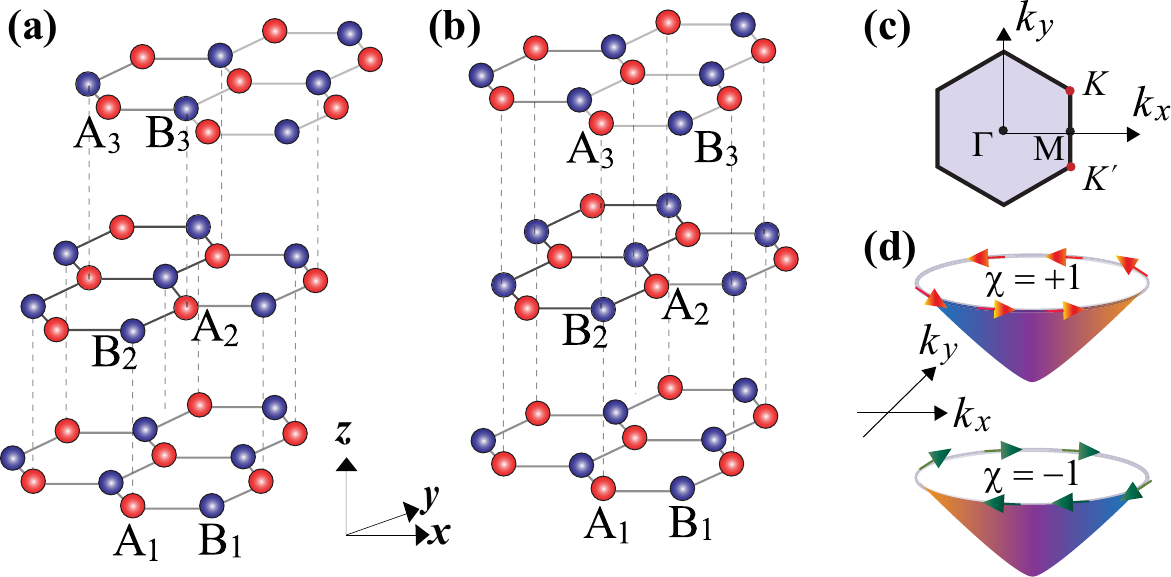}} \caption{Real-space structures of TLG with (a) ABC stacking and (b) ABA stacking in Cartesian coordinates $(x,y,z)$. A$_\ell$ and B$_\ell$ label two sublattices on the $\ell$-th layer, and the dotted lines connect the atoms that are exactly aligned in the $z$ direction. (c) Brillouin zone. $K$, $K'$, $\Gamma$ and $M$ denote four high-symmetry points. (d) Schematic diagram of the usual Rashba-induced band spin chirality $\chi=\pm1$ \cite{Zhai2014,Rashba}.}
\end{figure}

Here, we systematically study the Rashba spin splitting in TLG by establishing a general Hamiltonian with layer-dependent Rashba interaction induced by interface engineering \cite{ManKoo,ZolGmi2020,DayRay,AvsOch} from the top or bottom sides of TLG. We show that the main factors affecting the band spin splitting are the layer distribution of Rashba SOC and the stacking order of TLG. In detail, the more layers the Rashba SOC of the same sign and magnitude exist, the larger the band splitting is. For a pair of Rashba SOCs with the opposite sign but the same magnitude, we find that no spin splitting happens in ABC-TLG since the inversion symmetry is not broken (resulting in the complete cancellation of contributions from the opposite layers), whereas spin splitting appears for ABA-TLG due to its own lack of inversion symmetry. We also illustrate that gate voltage can modulate the spin polarization by breaking the spin or energy degeneracy near the band edges. Furthermore, we use density functional theory (DFT) calculations \cite{KriGol,zhai2022} to confirm the phenomena of Rashba spin splitting in TLG interfaced by Au layer(s), for which Rashba SOC and gate voltage work together.

Compared with Rashba bilayers \cite{zhai2022}, the Rashba trilayers we consider here have an additional middle layer that can be free from Rashba interaction and separates the top-bottom Rashba layers by a greater distance. Significantly, our results for TLG here reveal the nontrivial role of layer and symmetry in manipulating spin, applicable to not only TLG but also multilayers or other van der Waals materials. In addition, our results based on Rashba bilayers \cite{zhai2022} and trilayers reveal that an observable Rashba splitting can be induced without need of strictly controlling the number of graphene layers and the distribution uniformity between Rashba layers, facilitating wider applications of carbon films in spintronics.

We organize this article as follows. In Sec.~II, we establish a general model Hamiltonian for Rashba trilayers within the framework of tight-binding approximation. In Sec.~III, we present the results and discussion from the model. In Sec.~IV, we show the concrete examples of trilayer Rashba TLG from DFT calculations. In Sec.~V, other possible Rashba TLGs and device application are discussed. The last section is devoted to conclusion.

\begin{figure*}
\centerline{\includegraphics[width=17cm]{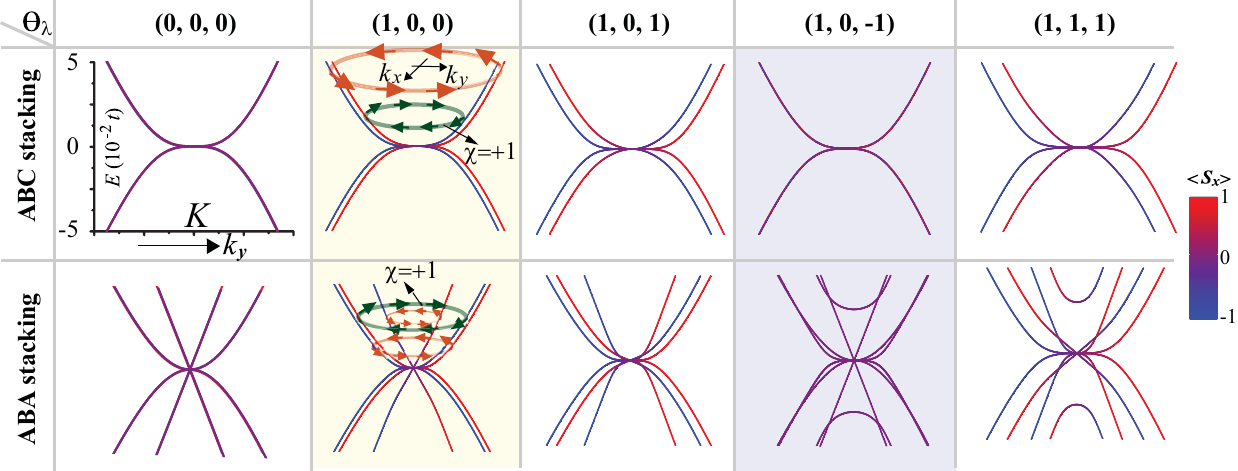}} \caption{Low-energy band structures for TLG with ABC stacking (top row) and ABA stacking (bottom row) near the $K$ point without gating ($U=0$). The momentum on the horizontal axis is along the $k_y$ direction, and the color indicates the spin-component average $\langle s_x\rangle$. We use stacking order to differentiate the rows and $\Theta_\lambda$ to differentiate the columns as $\lambda=7.14\times10^{-3}t$ is fixed. Specifically for the second column, we use the loop marked by arrow to denote the spin handness in the $(k_x,k_y)$ plane [$\chi=+1$ is labelled according to Fig.~1(d)], and other cases can be similarly judged. By contrast, the first column shows the Rashba-absent case, and the last column shows the trilayer-uniform Rashba case, whereas we focus beyond the first and last columns. All the bands are plotted with the same coordinate ranges as the first plot.}
\end{figure*}

\section{General Model Hamiltonian}

On basis of the tight-binding approximation \cite{Rashba,KanMe}, we begin with an empirical lattice Hamiltonian
\begin{equation}\label{TBH}
\begin{split}
{\cal H}=&-t\sum_{{{\langle
i,j\rangle}_{\ell}}\alpha}c_{\ell i,\alpha}^\dag
c_{\ell j,\alpha}-\sum_{{{\ell
i,\ell' j,}}\alpha}^{\ell\neq\ell'}\gamma_{ij} c_{\ell i,\alpha}^\dag c_{\ell j,\alpha}\\
&+\frac{i}{3}\sum_{\langle i,j\rangle_{\ell}\alpha\beta}
\lambda_\ell c_{\ell i,\alpha}^\dag({\bm s}\times\hat{{\bm d}}_{\ell i,\ell j})_{\alpha\beta}^zc_{\ell j,\beta}\\
&+\sum_{\ell i,\alpha}U_\ell c_{\ell i,\alpha}^\dag c_{\ell i,\alpha}+\sum_{i\alpha}\Delta_ic_{i\alpha}^\dag c_{i\alpha},
\end{split}
\end{equation}
where $c_{\ell i,\alpha}^\dag$ creates an electron with spin polarization $\alpha$ at site $i$ on layer $\ell$ ($\ell=1,2,3$ for bottom layer, middle layer and top layer, respectively), $\langle i,j\rangle$ runs over all the nearest-neighbor-hopping sites, $\alpha~(\beta)$ denotes spin up (down), ${\bm s}=(s_x,s_y,s_z)$ is the spin Pauli operator, $\hat{{\bm d}}_{ij}$ is the unit vector pointing from site $i$ to site $j$. The parameter $t$ describes the intralayer nearest-neighbor hopping energy, $\lambda_\ell$ is the $\ell$-dependent Rashba spin-orbit energy, $\gamma_{\langle i,j\rangle}$ denotes the interlayer coupling energy between site $i$ on layer $\ell$ and site $j$ on layer $\ell'$ ($\ell'\neq\ell$), $U_\ell$ indicates the effective layer-dependent potential induced by gate voltage, and $\Delta_i$ denotes the on-site energy. The hopping parameter $t$ satisfies $\upsilon=\sqrt3at/2\hbar$, where $a=2.46~{\AA}$ is the lattice constant and $\hbar$ is the reduced Planck constant. For simplicity, we here mainly consider the interlayer nearest-neighbor interaction ($\gamma_{\langle i,j\rangle}=\gamma$), which roughly captures the main band features, while other longer-distance interlayer interactions are useful to modify the band edges \cite{JhaCra}.

Corresponding to the five terms in Eq.~(\ref{TBH}) in order, we take $\psi=(\psi_{\ell=1},\psi_{\ell=2},\psi_{\ell=3})$ with $\psi_{\ell}=\{\psi_{A_\ell\uparrow},\psi_{A_\ell\downarrow},\psi_{B_\ell\uparrow},
\psi_{B_\ell\downarrow}\}$ as the atomic basis set and perform the Fourier transformation \cite{CasNet,KanMe}, and then obtain the effective momentum-space Hamiltonian near the Fermi energy as
\begin{equation}\label{H(p)}
H(\bm p)=H_\upsilon(\bm p)+H_\gamma+H_\lambda+H_U+H_\Delta,
\end{equation}
where the last term $H_\Delta$ (nearly zero for bare TLG \cite{ZolGmi2022}) depends on the interface details (see the example in Fig.~5, while we do not write specific expressions here) and contributes only to the diagonal term of the Hamiltonian, and the other terms are written as
\begin{equation}\label{Hp_part}
\begin{split}
H_\upsilon=&\upsilon{\bm I_\tau}(\sigma_xp_x+\xi\sigma_yp_y){\bm I_s},\\
H_\gamma=&\frac{\gamma}{2}(\tau_x\sigma_x-\tau_y\sigma_y){\bm I_s},\\
H_\lambda=&\frac{\lambda}{2}\tau_{z,\phi}(\sigma_xs_y-\xi\sigma_ys_x),\\
H_U=&U\tau_{z,\varphi}{\bm I_\sigma}{\bm I_s},
\end{split}
\end{equation}
where ${\bm p}=(p_x, p_y)$ describes the momentum [$K$ ($K'$) as coordinate origin], $\xi=\pm1$ represents valley $K$ ($K'$), ${\bm s}=(s_x, s_y, s_z)$ is spin Pauli operator, ${\bm \sigma}=(\sigma_x, \sigma_y, \sigma_z)$ is intralayer sublattice pseudospin Pauli operator,  ${\bm I_{s(\sigma)}}$ labels the $2\times2$ identity matrix in the $\bm s~(\bm\sigma)$ space, ${\bm I_\tau}$ is the $3\times3$ identity matrix in the layer pseudospin $\bm \tau$ space. For each term in Hamiltonian~(\ref{Hp_part}), $H_\upsilon$ indicates the massless Dirac term ($\upsilon$ is the Fermi velocity in monolayer graphene), $H_\gamma$ denotes the interlayer nearest-neighbor coupling ($\gamma$ is the strength), $H_\lambda$ represents the Rashba SOC ($\lambda$ is the strength), and $H_U$ represents the effective potential from gate voltage ($2U$ corresponds to the vertical bias). In particular, to facilitate the description of the layer pseudospin of trilayer graphene, we here employ the $3\times3$ Pauli-like pseudospin matrices
\begin{equation*}
\tau_x\equiv
\left(\begin{array}{ccc}
0&1&0\\
1&0&1\\
0&1&0\\
\end{array}\right),
\tau_y\equiv
\left(\begin{array}{ccc}
0&-i&0\\
i&0&-\nu i\\
0&\nu i&0\\
\end{array}\right),
\tau_{z,\vartheta}\equiv
\left(\begin{array}{ccc}
\vartheta_1&0&0\\
0&\vartheta_2&0\\
0&0&\vartheta_3\\
\end{array}\right),
\end{equation*}
where $\vartheta$ denotes $\phi$ in $H_\lambda$ and $\varphi$ in $H_U$, and $\nu$ is $+1$ for the ABC-stacking case and $-1$ for the ABA-stacking case. To simplify the parameter description, we define $\Theta_\lambda=(\phi_1,\phi_2,\phi_3)$ and $\Theta_U=(\varphi_1,\varphi_2,\varphi_3)$, where it satisfies $U_\ell=U\phi_\ell$ and $\lambda_\ell=\lambda\varphi_\ell$ corresponding to Eq.~(\ref{TBH}). Without special instruction, a fixed value of $\lambda=7.14\times10^{-3}t$ is used below to perform model calculations, whereas the main conclusions (Abstract) do not depend on the choice of $\lambda$ value.

It should be noted that a recent work by Zollner {\it et al.} have provided an effective Hamiltonian that includes additional proximity exchange and spin-orbit couplings in TLG-based van der Waals heterostructures \cite{ZolGmi2022} in addition to our concerned Rashba terms. There, the Rashba term is a perturbation term and our-of-plane spin polarization is focused. In contrast, we concentrate on the Rashba problem with respect to band spin chirality [Fig.~1(c)]. The core issue we are concerned about is the layer-dependent Rashba SOC, not only the amplitude, but more importantly the sign.

\section{Results and discussion from the model}

We start this section with two instructions below. Firstly, we consider the situation of spin splitting near the valley $K$ here, while the situation near the other valley $K'$ can be achieved by time-reversal symmetry \cite{Zhai2014}. Secondly, we plot the energy bands along the $k_y$ direction and mainly calculate the average value of spin component $\langle s_x\rangle$, and schematically draw the chirality of spin vortex [as sketched in Fig.~1(d)] where appropriate. Note that it always satisfies $\sum_{i=x,y,z}|\langle s_i\rangle|^2=1$, where $\langle s_z\rangle$ is shown in Fig.~4 due to the necessity of analyzing the spin orientation. For simplicity, we here set $\hbar/2$---the unit of $s_i$ ---to be 1.

Without gate voltage, {\it i.e.} $U_\ell=0$ in model~(\ref{TBH}), we plot the low-energy band structures in Fig.~2 by using the Rashba parameter $\lambda=7.14\times10^{-3}t$. For $\Theta_\lambda=(1,0,0)$, $\Theta_\lambda=(1,0,1)$ and $\Theta_\lambda=(1,1,1)$, it is seen that the number of subbands is doubled compared with $\Theta_\lambda=(0,0,0)$, as expected from the traditional Rashba effect \cite{ManKoo}. Judging from the strength of band spin splitting, we find that the more layers the Rashba SOC of the same sign and magnitude exist, the larger the band splitting is, independent of the stacking order. From the effect of band splitting, the Rashba effect localized in each layer can be equivalently considered to be equally divided by three layers through interlayer van der Waals coupling, similar to the previous finding in BLG \cite{zhai2022}. For the particular case $\Theta_\lambda=(1,0,-1)$---the spatially-separated two Rashba SOCs of the same magnitude but the opposite sign, no spin splitting occurs in ABC-TLG due to the complete cancellation of contributions from the opposite layers, whereas nonzero spin splitting is observed for ABA-TLG. Essentially, inversion symmetry is preserved for ABC-TLG, while ABA-TLG itself has no inversion symmetry even without the Rashba term. Note that the situation of $\Theta_\lambda=(1,1,1)$ here is shown only for comparison, while we mainly focus on the situations of $\Theta_\lambda=(1,0,\pm1)$ where Rashba SOCs are induced by proximity effect from the top and bottom sides. As a result, the band spin splitting depends strongly on the layer-distribution and sign of Rashba SOC $(\Theta_\lambda$) as well as the ABA or ABC stacking order of TLG. In other words, symmetry and stacking-order play a significant role in determining whether Rashba splitting exsits in energy band. Notably, the symmetry breaking here refers to the global structure rather than the local structure (the top or bottom interface).

\begin{figure*}
\centerline{\includegraphics[width=17cm]{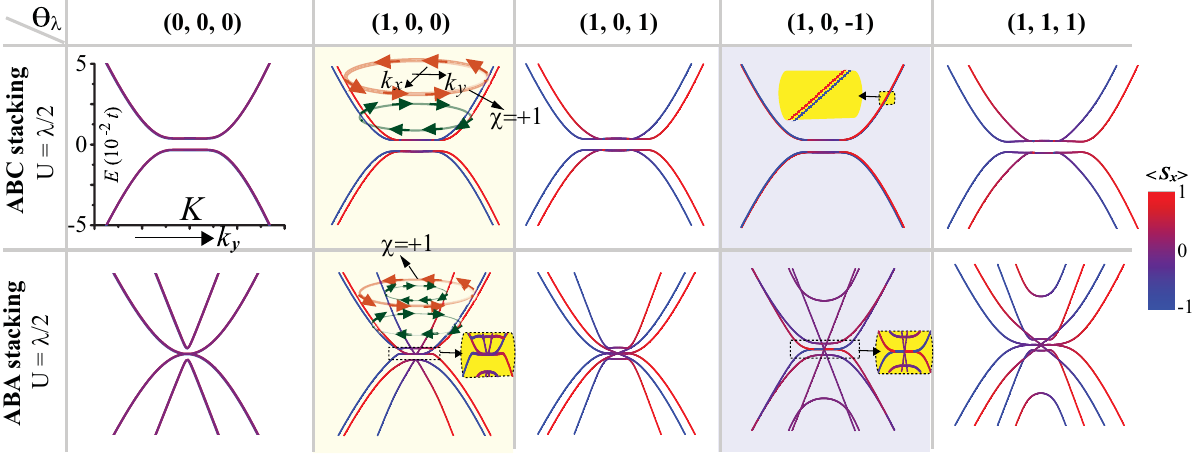}} \caption{Low-energy band structures for gated TLG. This figure is obtained by setting the gate voltage $2U=\lambda=7.14\times10^{-3}t$ and $\Theta_U=(1,0,-1))$ on basis of Fig.~2. The insets enlarge the local band regions.}
\end{figure*}

\begin{figure}
\centerline{\includegraphics[width=8.5cm]{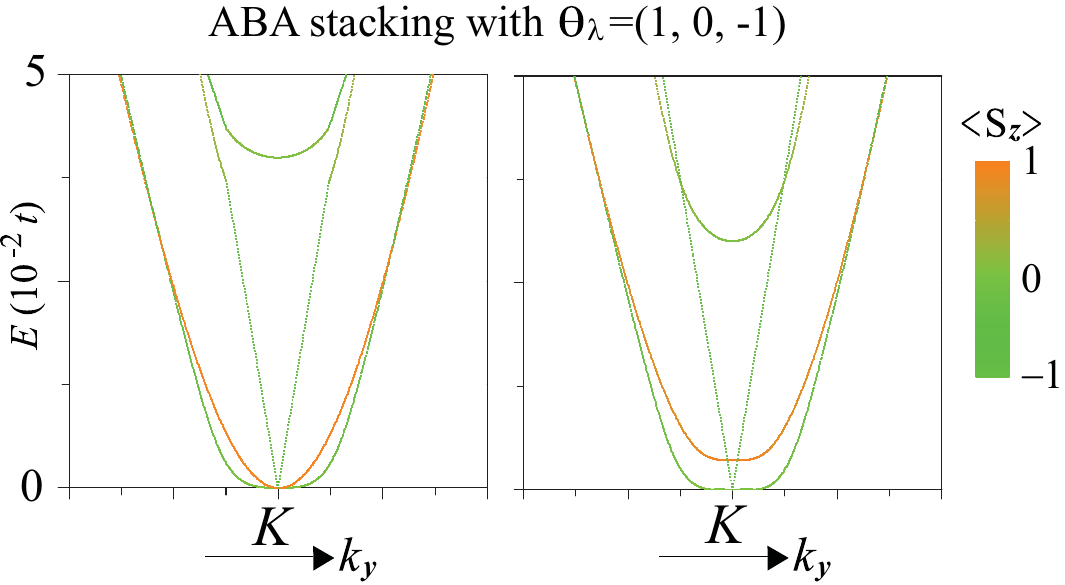}} \caption{The spin-component average $\langle s_z\rangle$ denoted by color in conduction bands for ABA-TLG with $\Theta_\lambda=(1,0,-1)$ by fixing $\lambda=7.14\times10^{-3}t$. The gate voltage reads $U=0$ (left panel) and $2U=7.14\times10^{-3}t$, $\Theta_U=(1,0,-1)$ (right panel).}
\end{figure}

Since we are mainly concerned with the electronic states near the Dirac point,
we further use the Green's function method \cite{ZhangSahu} to obtain the effective Hamiltonian for the simple case in Fig.~2, where the last two terms in Eq.~(\ref{TBH}) are absent. For the TLG with ABC-stacking, a four-band effective Hamiltonian in basis \{$\psi_{{B_1}\uparrow}$, $\psi_{{B_1}\downarrow}$, $\psi_{{A_3}\uparrow}$, $\psi_{{A_3}\downarrow}$\} is obtained as
\begin{equation}\label{band0_ABC}
\widetilde{H}_{\rm eff}^{\rm ABC}=\left[\frac{(\upsilon\pi^\dag)^3}{2\gamma^2}(\sigma_x+i\sigma_y){\bm I}_s
+\lambda\sum_{i=1}^3\phi_i\frac{(\upsilon\pi^\dag)^2}{\gamma^2}\sigma's'\right]
+h.c.,
\end{equation}
where $\pi=p_x-i\xi p_y$ is defined, $\Theta_\lambda=(\phi_1,\phi_2,\phi_3)$ has been defined below Eq.~(\ref{Hp_part}), and the orbit pseudospin operator $\sigma'$ and the spin operator $s'$ read, respectively,
\begin{equation}
\sigma'=\frac{1}{2}(-i\xi\sigma_x+\sigma_y),~s'=\frac{1}{2}(s_x+i\xi s_y).
\end{equation}
The term of $\sum_{i=1}^3\phi_i=0$ in Eq.~(\ref{band0_ABC}) agrees with the absence of spin splitting in Fig.~2 when inversion symmetry is not broken for $\Theta_\lambda=(1,0,-1)$. The magnitude of band spin splitting depends strongly on the value of $\sum_{i=1}^3\phi_i$, completely consistent with the calculations in Fig.~2.

For the TLG with ABA-stacking, an eight-band low-energy effective Hamiltonian in basis \{$(\psi_{{{\rm B}_1\uparrow}}-\psi_{{{\rm B}_3}\uparrow})/\sqrt2$, $(\psi_{{{\rm B}_1\downarrow}}-\psi_{{{\rm B}_3}\downarrow})/\sqrt2$,  $(\psi_{{{\rm A}_1\uparrow}}-\psi_{{{\rm A}_3}\uparrow})/\sqrt2$, $(\psi_{{{\rm A}_1\downarrow}}-\psi_{{{\rm A}_3}\downarrow})/\sqrt2$, $(\psi_{{{\rm B}_1\uparrow}}+\psi_{{{\rm B}_3}\uparrow})/\sqrt2$, $(\psi_{{{\rm B}_1\downarrow}}+\psi_{{{\rm B}_3}\downarrow})/\sqrt2$, $\psi_{{A_2}\uparrow}$, $\psi_{{A_2}\downarrow}$\} is obtained as
\begin{equation}
\tilde{H}_{\rm eff}^{\rm ABA}=
\left(\begin{array}{cc}
h_m&D\\
D^\dag&h_b\\
\end{array}\right),
\end{equation}
where the different items are expressed as
\begin{equation}
\begin{split}
&h_m=[\upsilon\pi^\dag(\sigma_x+i\sigma_y){\bm I}_s+h.c.]+\frac{\phi_1+\phi_3}{2}\lambda(\sigma_xs_y+\xi\sigma_ys_x),\\
&h_b=\left[-\frac{(\upsilon\pi^\dag)^2}{\sqrt2\gamma}(\sigma_x+i\sigma_y){\bm I}_s
+(\sum_{i=1}^3\phi_i+\phi_2)\frac{\upsilon\pi^\dag}{2\sqrt2\gamma}\sigma's'\right]+h.c.,\\
&D=\frac{\phi_3-\phi_1}{2}\xi\left[\frac{\upsilon\pi^\dag}{\sqrt2\gamma}\sigma's'-
(\sigma's')^\dag\right].
\end{split}
\end{equation}
Here, $h_m$ and $h_b$ describe the Rashba-modified monolayer-like and bilayer-like Hamiltonian \cite{CasNet}, respectively; $D$ denotes the coupling between $h_m$ and $h_b$. For the case of $\Theta_\lambda=(\phi_1,\phi_2,\phi_3)=(1,0,-1)$, we have $\phi_1+\phi_3=0$, $\sum_{i=1}^3\phi_i=0$ and $\phi_2=0$, and therefore $h_m$ and $h_b$ return to the original forms of massless Dirac and massive parabolic dispersion \cite{BaoJing}, whereas the nonzero $D$ modifies the Hamiltonian and induced Rashba splitting.

Then, we show the influence of gate voltage on band structure in Fig.~3 by fixing $2U=\lambda=7.14\times10^{-3}t$ on basis of Fig.~2. For $\Theta_\lambda=(0,0,0)$, it is seen that a band gap is opened in ABC-TLG, but semi-metallic property is preserved in ABA-TLG in spite of a local energy gap opened for the linear-dispersion Dirac cone, consistent with previous results \cite{LuiLi,BaoJing}. For $\Theta_\lambda=(1,0,-1)$, spin degeneracy is opened even though it is very weak (similar to the high-order perturbation effect in bilayer graphene \cite{zhai2022}), because gate voltage breaks inversion symmetry in ABC-TLG.

\begin{figure*}
\centerline{\includegraphics[width=17cm]{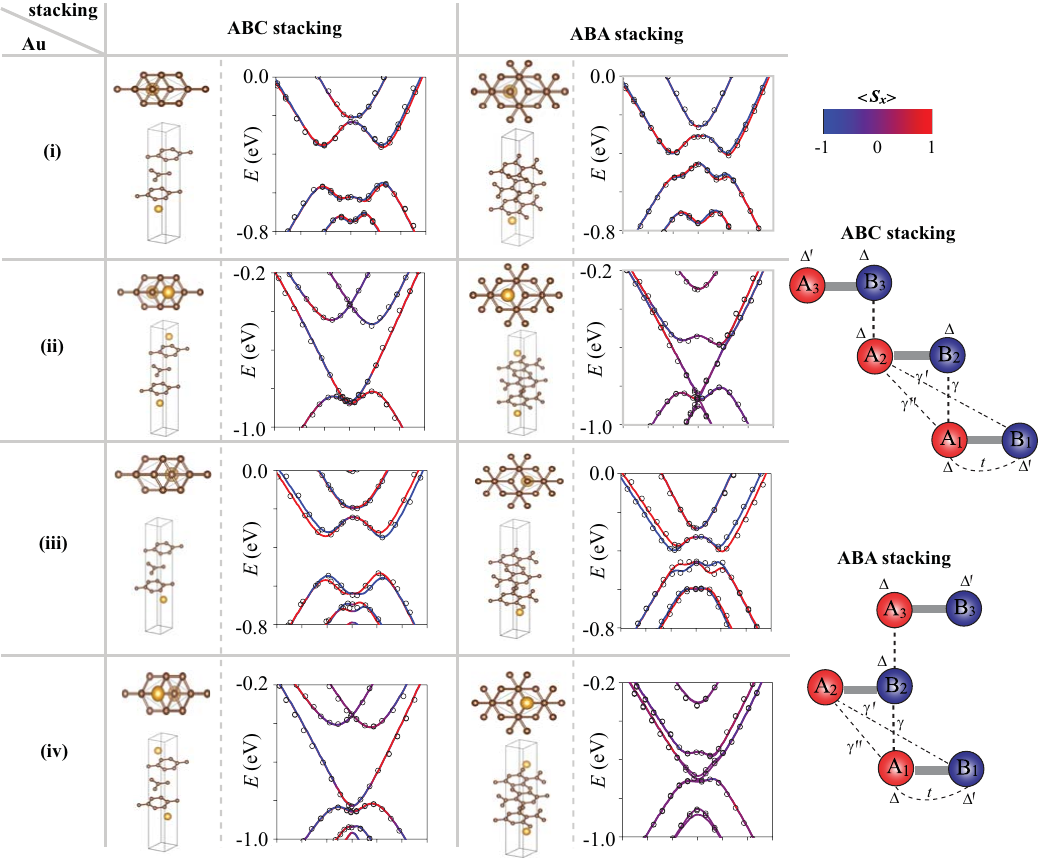}} \caption{Band structures calculated by DFT (solid lines) and fitted by lattice model (circles) for TLG interfaced by Au layer(s). The unit cell is sketched from the top and side views for each structure, where we mainly consider the good-symmetry cases of Au atoms located at the hexagonal centers [case (i) and case (ii)] or hexagonal vertices [case (iii) and case (iv)] of their nearest-neighboring graphene layer. The right insets show the side views of ABC- and ABA-TLG, where site energies $\Delta$, $\Delta'$ and hopping parameters $t$, $\gamma$, $\gamma'$, $\gamma''$ are denoted. See Table~II for complete fit parameters.}
\end{figure*}

Regardless of spin degeneracy in Fig.~2 and Fig.~3, we notice that the spin-component average satisfies $\langle s_x\rangle\simeq0$ in the vicinity of valley $K$ (further calculations not shown here indicate that $\langle s_y\rangle\simeq0$ and $|\langle s_z\rangle|\simeq1$). This means that spin flipping happens when the electronic states go across valley $K$. For ABA-TLG with $\Theta_\lambda=(1,0,-1)$, it is shown that $\langle s_x\rangle\simeq0$ (actually $\langle s_y\rangle\simeq0$) holds in Fig.~2 for $U=0$, and becomes nonzero for $U\neq0$ in Fig.~3. To clarify the spin orientation for $\Theta_\lambda=(1,0,-1)$, we further plot the spin-component average $\langle s_z\rangle$ of electronic states in conduction bands in Fig.~4. We find that $\langle s_z\rangle\simeq1$ holds, reflecting the spin signature dominated by the out-of-plane orientation in the situation of layered-opposite Rashba SOC \cite{zhai2022}.

In addition, we sketch the band spin chirality by taking $\Theta_\lambda=(1,0,0)$ as an example in Fig.~2 and Fig.~3 when the electron energy deviates from the neutral point ($E=0$). It is seen that there is one subband with $\chi=+1$ and one subband with $\chi=-1$ for ABC-TLG, whereas the subband number with $\chi=-1$ is one more than that with $\chi=+1$ for ABA-TLG. It is expected that this difference should be identifiable from the spin Hall experiment \cite{GhiKav,BenTor}. The analysis of band spin chirality here is also applicable to other situations in Fig.~2 and Fig.~3.

\section{Trilayer Rashba TLG from DFT calculations}

\begin{table*} \label{T1}\footnotesize
\centering
\caption{Fitting parameters for the DFT calculations in Fig.~5. The parameters $t$, $\gamma$, $\gamma'$ and $\gamma''$ are the hopping energies, $\Delta$ and $\Delta$' denote the site energies, $\mu$ denotes the chemical potential induced by doping from Au layer(s), as sketched in the right insets of Fig.~5. The Rashba and bias parameters read $(\lambda,\Theta_\lambda)$ and $(U,\Theta_U)$, respectively, as shown in model~(\ref{H(p)}).}
\begin{threeparttable}
\begin{tabular}{rrrrrrrrrcr}
\toprule[1pt]
\toprule[1pt]
\quad \quad \quad System&\quad{{\rm ABC}-(i)}&\quad\quad{{\rm ABC}-(ii)}&\quad\quad{{\rm ABC}-(iii)} &\quad\quad{{\rm ABC}-(iv)}&\quad\quad{{\rm ABA}-(i)}&\quad\quad{{\rm ABA}-(ii)}&\quad\quad{{\rm ABA}-(iii)}&\quad\quad{{\rm ABA}-(iv)}&\quad\quad\\
\hline
\quad\quad $t$ (eV)\quad &\quad2.572&\quad\quad2.386&\quad\quad2.569&\quad\quad2.423 &\quad\quad 2.576&\quad\quad 2.489&\quad\quad2.528 &\quad\quad2.461\\
\quad\quad $\gamma$ (eV)\quad&\quad0.325&\quad\quad0.384&\quad\quad0.369 &\quad\quad 0.385 &\quad\quad0.321 &\quad\quad0.275 &\quad\quad0.322 &\quad\quad0.221\\
\quad\quad $\gamma'$ (eV)&\quad\quad-0.192&\quad\quad-0.216&\quad\quad-0.237 &\quad\quad -0.279 &\quad\quad-0.183 &\quad\quad-0.212 &\quad\quad-0.229&\quad\quad-0.192\\
\quad\quad $\gamma''$ (eV)\quad&\quad-0.144&\quad\quad-0.167&\quad\quad-0.158 &\quad\quad -0.164 &\quad\quad-0.148 &\quad\quad-0.165 &\quad\quad-0.159&\quad\quad-0.167\\
\quad\quad $\Delta$ (eV)\quad &\quad-0.050&\quad0.086&\quad0.076&\quad\quad0.086 &\quad\quad0 &\quad\quad0 &\quad\quad0.168 &\quad\quad-0.160\\
\quad\quad $\Delta '$ (eV)\quad &\quad0.027&\quad\quad-0.023&\quad0.083&\quad\quad0 &\quad\quad-0.082 &\quad\quad0 &\quad\quad0.135 &\quad\quad0\\
\quad\quad $\lambda$ (meV)  &\quad\quad43.4&\quad\quad43.2&\quad\quad93.6&\quad\quad93.3 &\quad\quad 43.2&\quad\quad 42.9&\quad\quad 93.2&\quad\quad92.8\\
\quad\quad $\Theta_\lambda$\quad &\quad\quad(1,0,0)&\quad\quad(1,0,-1)&\quad(1,0,0)&\quad\quad(1,0,-1) &\quad\quad(1,0,0) &\quad\quad(1,0,-1) &\quad\quad(1,0,0) &\quad\quad(1,0,-1)\\
\quad\quad $U$ (eV)\quad &\quad-0.261 &\quad\quad-0.003&\quad\quad-0.202&\quad\quad-0.003 &\quad\quad-0.281 &\quad\quad-0.305 &\quad\quad-0.258 &\quad\quad-0.301\\
\quad\quad $\Theta_U$ &\quad\quad(1,-0.75,-1)&\quad\quad(0,1,0)&\quad(1,0.25,-1)&\quad\quad(0,1,0) &\quad\quad(1,-0.86,-1) &\quad\quad(0,1,0) &\quad\quad(1,0.67,-1) &\quad\quad(0,1,0)\\
\quad\quad $\mu$ (eV)\quad &\quad-0.505 &\quad\quad-0.832&\quad\quad-0.502&\quad\quad-0.861 &\quad\quad-0.494 &\quad\quad-0.865 &\quad\quad-0.616 &\quad\quad-0.702\\
\bottomrule[1.0pt]
\end{tabular}
\end{threeparttable}
\end{table*}

By employing DFT calculations, we verify the Rashba spin splitting in model Hamiltonian~(\ref{TBH}) by contacting TLG with Au layer(s) from the top or bottom side(s), as an example but not limited to this example [see other candidates in Table~(\ref{T1})]. All the calculations are carried out using the VASP package \cite{Kres} based on the first-principles plane wave pseudopotential method and density functional theory.  The exchange correlation functional adopts the Perdew-BurkeErnzerhof (PBE) \cite{PerRu}, the plane wave basis energy cutoff is set as 500~eV and the $k$-points setup uses the Monkhorst-Pack scheme, with 15$\times$15$\times$1 $k$-point meshes. Geometric optimization is performed by electron relaxation, where the convergence value of energy is set as $10^{-6}$~eV and the force of all atoms is less than 0.01~eV/nm. The unit cell has six C atoms and one (two) Au atom(s), and a vacuum region of 20~$\AA$ is used to avoid the interactions between neighboring slabs. The lattice parameter of graphene is a = 2.46~$\AA$, with TLG interlayer distances of $d =3.3~\AA$ \cite{JhaCra}. To better observe the Rashba splitting effect, we artificially set the spacing between Au atom and graphene layer at set as 2.7~$\AA$ (smaller than the equilibrium position $3.2-3.4$~$\AA$). Undoubtedly, the interlayer distance obtained from different experimental samples may vary \cite{KriGol}, but the nature of Rashba induced band splitting will not be affected.

In Fig.~5, we show the DFT band structures for TLG in proximity with Au layer(s), where case (i) and case (iii) [case (ii) and case (iv)] are the single (dual)-interface situations, with the difference that cases (i) and case (ii) [case (ii) and case (iv)] correspond to the situation that Au atoms locate at the hexagonal centers (vertices) of their nearest-neighboring graphene layer. Our results indicate that the Rashba splitting indeed depends strongly on the stacking order and structural symmetry. For the same stacking order, case (iii) has larger splitting than case (i), as expected from Fig.~2. No spin splitting happens when inversion symmetry is preserved for case (ii) of ABC-TLG, likewise for case (iv) of ABC-TLG.
To demonstrate the spin orientation, we also show $\langle s_x\rangle$ by color in Fig.~5. It is worth nothing that $\langle s_x\rangle$ for case (ii) and case (iv) of ABC-TLG is zero (opposite spin is mixed, resulting in spin degeneracy).

Furthermore, we fit the DFT results by the tight-binding model parameters, as listed in Table~II. It is judged that the terms of Rashba SOC and gate voltage in model~(\ref{TBH}) are simultaneously induced. The Rashba term originates from the hopping between 5$d$ orbital in Au atom and $\pi$ orbital in carbon atom \cite{KriGol}. The chemical potential $\mu<0$ holds because there is electron transfer from Au atoms to carbon atoms. This charge transfer happens between the interlayer nearest-neighboring orbital hybridization of Au and carbon atoms. It is the potential energy difference between TLG layers that gives the origin of the effective gate-voltage term.
Matching with the DFT data and tight-binding model, a more accurate effective Hamiltonian with a more extended form of orbital part is given as follows. For ABC stacking, we have
\begin{equation}\label{H_ABC}
\begin{split}
&H_{\gamma'}^{\rm ABC}=\upsilon'(\tau_+\sigma_-\pi+\tau_-\sigma_+\pi^\dag),\\
&H_{\gamma''}^{\rm ABC}=\upsilon''(\tau_+\pi^\dag+\tau_-\pi),\\
&H_\Delta^{\rm ABC}=\tau_1\otimes\left(\begin{array}{cc}
\Delta&0\\
0&\Delta'\\
\end{array}\right)+\Delta\tau_2\otimes {\bm I}_{\sigma}+\tau_3\otimes\left(\begin{array}{cc}
\Delta'&0\\
0&\Delta\\
\end{array}\right),\\
\end{split}
\end{equation}
where $\upsilon'=\sqrt3\gamma't/(2\hbar)$, $\upsilon''=\sqrt3\gamma''t/(2\hbar)$ with $a=2.46~{\AA}$ (see Fig.~5 for hopping parameters $t$, $\gamma'$ and $\gamma''$), $\sigma_\pm=(\sigma_x\pm i\sigma_y)/2$, $\tau_\pm=(\tau_x\pm i\tau_y^{\nu=1})/2$.
For ABC stacking, we have
\begin{equation}\label{H_ABA}
\begin{split}
&H_{\gamma'}^{\rm ABA}=\upsilon'(\tau_4\sigma_-\pi+\tau_5\sigma_+\pi^\dag)+h.c.,\\
&H_{\gamma''}^{\rm ABA}=\upsilon''(\tau_4\pi^\dag+\tau_5\pi)+h.c.,\\
&H_\Delta^{\rm ABA}=(\tau_1+\tau_3)\otimes\left(\begin{array}{cc}
\Delta&0\\
0&\Delta'\\
\end{array}\right)+\frac{\Delta}{2}\tau_2\otimes({\bm I}_{\sigma}-\sigma_z),
\end{split}
\end{equation}
For the layer-related $3\times3$ matrices $\tau_{i}~(i=1\rightarrow5)$ in Eq.~(\ref{H_ABC}) and Eq.~(\ref{H_ABA}), all non-zero matrix elements are listed below
\begin{equation}
\begin{split}
(\tau_1)_{11}=1,~
(\tau_2)_{22}=1,~
(\tau_3)_{33}=1,~
(\tau_4)_{12}=1,~
(\tau_5)_{23}=1,
\end{split}
\end{equation}
which are used to shorten the writing of formulas.

Besides the situations where Au atoms locate at the specific positions in Fig.~5, the model~(\ref{TBH}) that we build should be applicable in the other situations, for which the values of $(\lambda,\Theta_\lambda)$, $(U,\Theta_U)$ and $(\Delta,\Delta')$ should be modified during model fitting. Conceivably, when magnetic interactions are further considered in complex interfaces \cite{zhai2022,ZolGmi2020,SieFab}, more physics related to magnetism and topology can be achieved.

From the perspective of experimental feasibility, the freestanding form of the structures in Fig.~5 is still a challenge although is is theoretically possible. The existing reports indicate that graphene can indeed be embedded by Au atoms arranged in a monolayer structure, while additional substrate like Ni \cite{KriGol} or Co \cite{Ryb2018} is needed to avoid the tendency of Au atoms to arrange in clusters. In actual operations, one can use SiO$_2$ \cite{MarVar2016}, instead of Ni or Co, as the substrate or use bulk Au \cite{LeiTes2014} (viewed equivalently, due to the short-range nature of the proximity effect, only the surface Au monolayer works) to preserve the well-separated graphene states in Fig.~5 without the extra substrate-induced overlapping or hybridization effects.

\section{Other Possible Rashba TLGs and Device Application}

\begin{figure}
\centerline{\includegraphics[width=8.5cm]{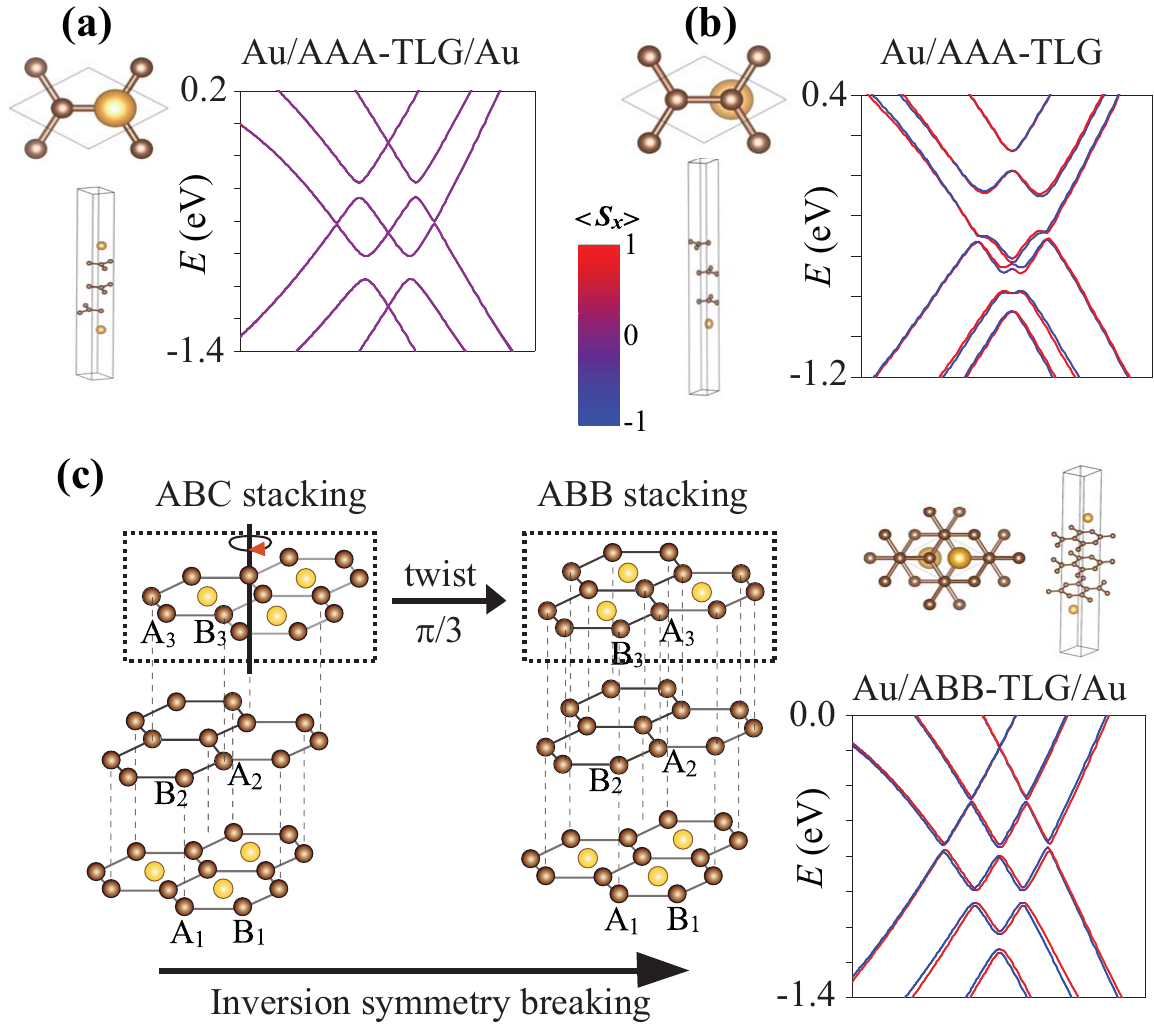}} \caption{Real-space structures and DFT band structures for Au/AAA-TLG in (a), Au/AAA-TLG/Au in (b) and Au/ABB-TLG/Au in (c). Specifically, the ABB-stacking TLG in (c) can be seen as the outcome of the ABC-stacking TLG after $\pi/3$ rotation of the top layer around the axis depicted in the figure while the other two layers remain unchanged.}
\end{figure}

We notice that the experimental observation has shown that there is an additional AAA-stacking TLG configuration (simple hexagonal stacking) \cite{BaoYao}, for which the electronic structure differ significantly from ABA-stacking and ABC-stacking. With this in mind, we perform further DFT calculations for Au/AAA-TLG in Fig.~6(a) and Au/AAA-TLG/Au in Fig.~6(b). The inversion symmetry is broken for the former but holds for the latter. As a result, no spin splitting is observed for Au/AAA-TLG/Au [$\Theta_\lambda=(1,0,-1)$], while spin splitting occurs for Au/AAA-TLG [$\Theta_\lambda=(1,0,0)$].

Actually, many structural configurations different from ABA and ABC stacking for TLG can occur under artificial control, typically twisted-angle control \cite{LiZhang2022,DevLed}. A notable recent theoretical work by Avisha {\it et. al.} has solved the problem of twisted TLG with Rashba SOC \cite{AviBand}, where the Rashba trilayer Moir\'{e} lattice especially at small twisted angle ($\theta$ is near $1^\circ$) is discussed, and as the twisted angle $\theta$ changes, the nonanalytic electronic charge density and the equilibrium spin currents are illustrated. Imaginably, the twist angle can easily alter or break the symmetry of untwisted parent structure. Based on these considerations, we choose a simpler case of twist-angle $\pi/3$ to discuss the problem, as shown in Fig.~6(c), where ABC stacking with inversion symmetry can become ABB stacking without inversion symmetry after the top layer is rotated. Notably, the TLG with ABB stacking does exist in the twist angle experimental samples \cite{LiZhang2022,DevLed}. As expected, spin splitting happens for ABB stacking here, in stack contrast to ABC stacking (Fig.~5). Undoubtedly, Moir\'{e} patterns and flat bands under small twist angle should appear \cite{LiZhang2022,DevLed,AviBand}, but the Rashba splitting mechanism dictated by symmetry breaking should not be altered.

\begin{figure}
\centerline{\includegraphics[width=8.5cm]{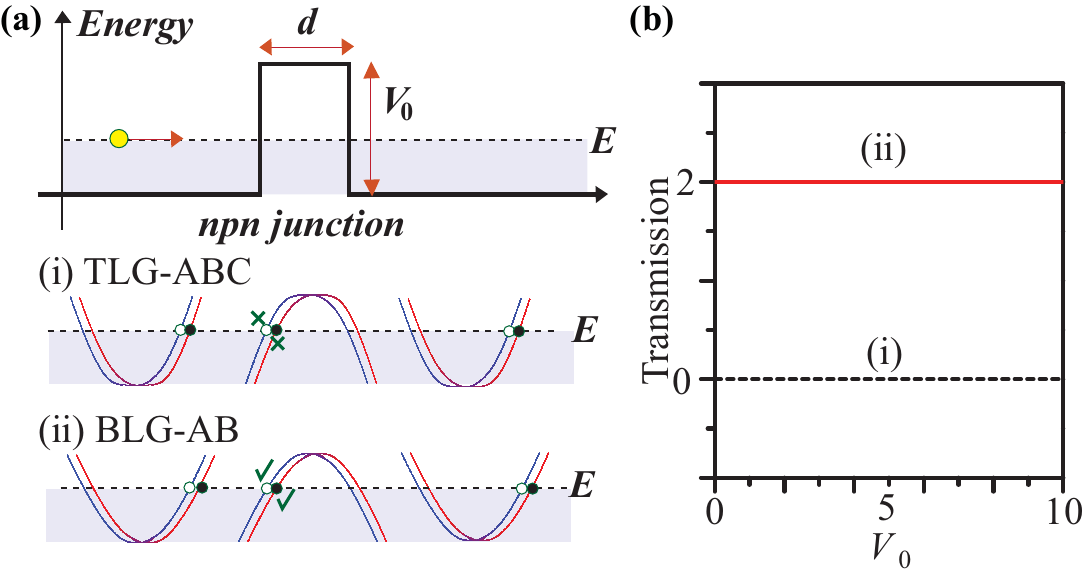}} \caption{(a) Quantum tunnelling in an $npn$ junction (top) for (i) Rashba TLG with ABC-stacking (middle) and (ii) Rashba BLG with AB stacking (bottom). $d$ is the barrier width, $E$ represents the energy of incident electron, and $V_0$ (in units of $E/|e|$) denotes the barrier height. For case (i) and case (ii), the dispersion is plotted in each region, and the solid and hollow points label two different electronic states at normal incidence in transport. The symbol $\times$ ($\surd$) indicates that the normally-incident electrons from the left are forbidden (allowed) from crossing the barrier. (b) Dependence of transmission on barrier height at normal incidence for two cases in (a). The parameters $E=20$~meV and $d=50$~nm are fixed.}
\end{figure}

Moreover, it is necessary to discuss the difference between Rashba trilayer and Rashba bilayer in quantum transport from the perspective of device application. For simplicity, we here discuss the situation of low-energy effective models based on the analytical expressions in Sec.~III. An $npn$ junction is proposed in Fig.~7(a). As analyzed by the band-to-band tunnelling, the transport should be forbidden at normal incidence for TLG with ABC stacking because the incident and transmitted (scattering region) states are orthogonal \cite{Zhai2014}, resulting in zero transmission. For BLG with AB stacking, the complete transmission should happen at normal incidence because the incident and reflected states are orthogonal, resulting in Klein tunneling \cite{CasNet}. As expected, the calculated transmission in Fig.~7(b) is independent of barrier height, holding for any $\Sigma_i\phi_i\neq0$ in our calculations.

Essentially, the results in Fig.~7 are a manifestation of band topology characterized by the Berry's (geometric) phase $\Phi_{\rm B}$, defined by $\Phi_{\rm B}=\oint_Cd{\bm p}\cdot{\cal A}(\bm p)$, where ${\cal A}(\bm p)=\langle\psi(\bm p)|i{\bm \nabla}_{\bm p}|\psi(\bm p)\rangle$ is the Berry connection for the wave function $\psi(\bm p)$. Using the effective Hamiltonian in Eqs.~(4)-(5) for Rashba TLG with ABC stacking, one can prove that $\Phi_{\rm B}$ changes from $3\pi$ under $\Sigma_i\phi_i=0$ to $2\pi$ under $\Sigma_i\phi_i\neq0$. In comparison, $\Phi_{\rm B}$ changes from $2\pi$ under $\Sigma_i\phi_i=0$ to $\pi$ under $\Sigma_i\phi_i\neq0$ for Rashba BLG \cite{Zhai2014}. This means, to observe Klein tunneling, $\Phi_{\rm B}/\pi$ should be an odd number \cite{CasNet,Zhai2014,zhai2022}, which also applies to the situation of AAA or ABB stacking as well as the complex situation of TLG with ABA stacking when both the Dirac-like dispersion and the parabolic-like dispersion coexist.

\section{Conclusions}

By adding the layer-dependent Rashba interaction to the usual Hamiltonian of trilayer graphene, we have got four useful bits of information. First, the situation of band spin splitting in Rashba TLG is determined by the layer-distribution and sign of Rashba interaction and especially the stacking order. Second, the more layers the Rashba SOC of the same sign and magnitude exist, the larger the band splitting is. Third, for the spatially-separated two Rashba SOCs of the opposite sign but the same magnitude, no spin splitting happens in ABC-TLG with inversion symmetry while obviously weakened spin splitting is observed for ABA-TLG without inversion symmetry. Forth, gate voltage is helpful to modulate the spin splitting by opening the energy gaps in the vicinity of energy valleys. To confirm the phenomena of Rashba spin splitting,  we have used DFT calculations in the platform of TLG interfaced by Au layer(s), which induce simultaneously the terms of Rashba SOC and gate voltage in TLG. Our results reveal the significance of layer and symmetry of TLG in manipulating spin and can be extended to other multilayer graphene or van der Waals quantum materials.
\\

This work was supported by the NSFC with Grant No.~62374088 and No.~12074193.


\begin{thebibliography}{99}%
\makeatletter
\providecommand \@ifxundefined [1]{%
 \@ifx{#1\undefined}
}%
\providecommand \@ifnum [1]{%
 \ifnum #1\expandafter \@firstoftwo
 \else \expandafter \@secondoftwo
 \fi
}%
\providecommand \@ifx [1]{%
 \ifx #1\expandafter \@firstoftwo
 \else \expandafter \@secondoftwo
 \fi
}%
\providecommand \natexlab [1]{#1}%
\providecommand \enquote  [1]{``#1''}%
\providecommand \bibnamefont  [1]{#1}%
\providecommand \bibfnamefont [1]{#1}%
\providecommand \citenamefont [1]{#1}%
\providecommand \href@noop [0]{\@secondoftwo}%
\providecommand \href [0]{\begingroup \@sanitize@url \@href}%
\providecommand \@href[1]{\@@startlink{#1}\@@href}%
\providecommand \@@href[1]{\endgroup#1\@@endlink}%
\providecommand \@sanitize@url [0]{\catcode `\\12\catcode `\$12\catcode
  `\&12\catcode `\#12\catcode `\^12\catcode `\_12\catcode `\%12\relax}%
\providecommand \@@startlink[1]{}%
\providecommand \@@endlink[0]{}%
\providecommand \url  [0]{\begingroup\@sanitize@url \@url }%
\providecommand \@url [1]{\endgroup\@href {#1}{\urlprefix }}%
\providecommand \urlprefix  [0]{URL }%
\providecommand \Eprint [0]{\href }%
\providecommand \doibase [0]{http://dx.doi.org/}%
\providecommand \selectlanguage [0]{\@gobble}%
\providecommand \bibinfo  [0]{\@secondoftwo}%
\providecommand \bibfield  [0]{\@secondoftwo}%
\providecommand \translation [1]{[#1]}%
\providecommand \BibitemOpen [0]{}%
\providecommand \bibitemStop [0]{}%
\providecommand \bibitemNoStop [0]{.\EOS\space}%
\providecommand \EOS [0]{\spacefactor3000\relax}%
\providecommand \BibitemShut  [1]{\csname bibitem#1\endcsname}%
\let\auto@bib@innerbib\@empty

\bibitem{CasNet} A. H. Castro Neto, F. Guinea, N. M. R. Peres, K. S. Novoselov, and A. K. Geim, The electronic properties of graphene, Rev. Mod. Phys. {\bf 81}, 109 (2009).

\bibitem{KonGmi} S. Konschuh, M. Gmitra, and J. Fabian, Band-structure topologies of graphene: Spin-orbit coupling effects from first principles, Phys. Rev. B {\bf 82}, 245412 (2010).

\bibitem{SiPra} J. Sichau, M. Prada, T. Anlauf, T. J. Lyon, B. Bosnjak, L. Tiemann, and R. H. Blick, Resonance microwave measurements of an intrinsic spin-orbit coupling gap in graphene: A possible indication of a topological state, Phys. Rev. Lett. {\bf 122}, 046403 (2019).

\bibitem{NovFal} K. S. Novoselov, V. I. Fal'ko, L. Colombo, P. R. Gellert, M. G. Schwab, and K. Kim, A roadmap for graphene, Nature {\bf 490}, 192 (2012).

\bibitem{Yazyev} O. V. Yazyev, Emergence of magnetism in graphene materials and nanostructures, Rep. Prog. Phys. {\bf 73}, 056501 (2010).

\bibitem{PesMacs} D. Pesin and A. H. MacDonald, Spintronics and pseudospintronics in graphene and topological insulators, Nat. Mater. {\bf 11}, 409 (2012).

\bibitem{AvsTan} A. Avsar, J. Y. Tan, T. Taychatanapat, J. Balakrishnan, G. K. W. Koon, Y. Yeo, J. Lahiri, A. Carvalho, A. S. Rodin, E. C. T. O'Farrell, G. Eda, A. H. Castro Neto, and B. \"{O}zyilmaz, Spin-orbit proximity effect in graphene, Nat. Commun. {\bf 5}, 4875 (2014).

\bibitem{HanKaw} W. Han, R. K. Kawakami, M. Gmitra, and J. Fabian, Graphene spintronics, Nat. Nanotechnol. {\bf 9}, 794 (2014).

\bibitem{DayRay} J.-F. Dayen, S. J. Ray, O. Karis, I. J. Vera-Marun, and M. V. Kamalakar, Two-dimensional van der Waals spinterfaces and magnetic-interfaces, Appl. Phys. Rev. {\bf 7}, 011303 (2020).

\bibitem{ZolGmi2020} K. Zollner, M. Gmitra, and J. Fabian, Swapping exchange and spin-orbit coupling in 2D van der waals heterostructures, Phys. Rev. Lett. {\bf 125}, 196402 (2020).

\bibitem{AvsOch} A. Avsar, H. Ochoa, F. Guinea, B. \"{O}zyilmaz, B. J. van Wees, and I. J. Vera-Marun, Colloquium: Spintronics in graphene and other two-dimensional materials, Rev. Mod. Phys. {\bf 92}, 021003 (2020).

\bibitem{Magda} G. Z. Magda, X. Jin, I. Hagym\'{a}si, P. Vancs\'{o}, Z. Osv\'{a}th, P. Nemes-Incz\'{e}, C. Hwang, L. P. Bir\'{o}, and L. Tapaszt\'{o}, Room-temperature magnetic order on zigzag edges of narrow graphene nanoribbons, Nature. {\bf 514}, 608 (2014)

\bibitem{Slota} M. Slota, A. Keerthi, W. K. Myers, E. Tretyakov, M. Baumgarten, A. Keerthi, W. K. Myers, E. Tretyakov, M. Baumgarten, A. Ardavan, H. Sadeghi, C. J. Lambert, A. Narita, K. M\''{u}llen, and L. Bogani, Magnetic edge states and coherent manipulation of graphene nanoribbons, Nature. {\bf 557}, 691 (2018).

\bibitem{GonHer} H. Gonz\'{a}lez-Herrero, J. M. G\'{o}mez-Rodr\'{l}guez, P. Mallet, M. Moaied, J. Jose Palacios, C. Salgado, M. M. Ugeda, J.-Y. Veuillen, F. Yndurain, and I. Brihuega, Atomic-scale control of graphene magnetism using hydrogen atoms, Science {\bf 352}, 437 (2016).

\bibitem{GhiKav} T. S. Ghiasi, A. A. Kaverzin, A. H. Dismukes, D. K. de Wal, X. Roy, and B. J. van Wees, Electrical and thermal generation of spin currents by magnetic bilayer graphene, Nat. Nanotechnol. {\bf 16}, 788 (2021).

\bibitem{SieFab} J. F. Sierra, J. Fabian, R. K. Kawakami, S. Roche, and S. O. Valenzuela, Van der Waals heterostructures for spintronics and opto-spintronics, Nat. Nanotechnol. {\bf 16}, 856 (2021).

\bibitem{KanMe} C. L. Kane and E. J. Mele, $Z_2$ Topological Order and the Quantum Spin Hall Effect, Phys. Rev. Lett. {\bf 95}, 146802 (2005).

\bibitem{ManKoo} A. Manchon, H. C. Koo, J. Nitta, S. M. Frolov, and R. A. Duine, New perspectives for Rashba spin-orbit coupling, Nat. Mater. {\bf 14}, 871 (2015).

\bibitem{Hoque} A. M. Hoque, D. Khokhriakov, K. Zollner, B. Zhao, B. Karpiak, J. Fabian, and S. P. Dash, All-electrical creation and control of spin-galvanic signal in graphene and molybdenum ditelluride heterostructures at room temperature, Commun. Phys. {\bf 4}, 124 (2021).

\bibitem{Khokh} D. Khokhriakov, A. M. Hoque, B. Karpiak, and P. Dash, Gate-tunable spin-galvanic effect in graphene-topological insulator van der Waals heterostructures at room temperature, Nat. Commun. {\bf 11}, 3657 (2020).

\bibitem{TiwSri} P. Tiwari, S. K. Srivastav, S. Ray, T. Das, and A. Bid, Observation of time-reversal invariant helical edge-modes in bilayer graphene/WSe$_2$ heterostructure, ACS Nano {\bf 15}, 916 (2021).

\bibitem{DedFon} Y. S. Dedkov, M. Fonin, U. R\"{u}diger, and C. Laubschat, Rashba effect in the graphene/Ni(111) system, Phys. Rev. Lett. {\bf 100}, 107602 (2008).

\bibitem{Rashba} E. I. Rashba, Graphene with structure-induced spin-orbit coupling: Spin-polarized states, spin zero modes, and quantum Hall effect, Phys. Rev. B {\bf 79}, 161409(R) (2009).

\bibitem{MirSch} F. Mireles and J. Schliemann, Energy spectrum and Landau levels in bilayer graphene with spin-orbit interaction, New J. Phys. {\bf 14}, 093026 (2012).

\bibitem{MarVar2012} D. Marchenko, A. Varykhalov, M. R. Scholz, G. Bihlmayer, E. I. Rashba, A. Rybkin, A. M. Shikin, and O. Rader, Giant Rashba splitting in graphene due to hybridization with gold, Nat. Commun. {\bf 3}, 1232 (2012).

\bibitem{QiaoLi} Z. Qiao, X. Li, W.-K. Tse, H. Jiang, Y. Yao, and Q. Niu, Topological phases in gated bilayer graphene: Effects of Rashba spin-orbit coupling and exchange field, Phys. Rev. B {\bf 87}, 125405 (2013).

\bibitem{Zhai2014} X. Zhai and G. Jin, Reversing Berry phase and modulating Andreev reflection by Rashba spin-orbit coupling in graphene mono-and bilayers, Phys. Rev. B {\bf 89}, 085430 (2014).

\bibitem{WangKi} Z. Wang, D. K. Ki, H. Chen, H. Berger, A. H. MacDonald, and A. F. Morpurgo, Strong interface-induced spin-orbit interaction in graphene on WS$_2$, Nat. Commun. {\bf 6}, 8339 (2015).

\bibitem{MarVar2016} D. Marchenko, A. Varykhalov, J. S\'{a}nchez-Barriga, Th. Seyller, and O. Rader, Rashba splitting of 100 meV in Au-intercalated graphene on SiC, Appl. Phys. Lett. {\bf 108}, 172405 (2016).

\bibitem{FarTan} E. C. T. O'Farrell, J. Y. Tan, Y. Yeo, G. K. W. Koon, B. \"{O}zyilmaz, K. Watanabe, and T. Taniguchi, Rashba interaction and local magnetic moments in a graphene-BN heterostructure intercalated with Au, Phys. Rev. Lett. {\bf 117}, 076603 (2016).

\bibitem{KriGol} M. Krivenkov, E. Golias, D. Marchenko, J. S\'{a}nchez-Barriga, G. Bihlmayer, O. Rader, and A. Varykhalov, Nanostructural origin of giant Rashba effect in intercalated graphene, 2D Mater. {\bf 4}, 035010 (2017).

\bibitem{YangLoh} B. Yang, M. Lohmann, D. Barroso, I. Liao, Z. Lin, Y. Liu, L. Bartels, K. Watanabe, T. Taniguchi, and J. Shi, Strong electron-hole symmetric Rashba spin-orbit coupling in graphene/monolayer transition metal dichalcogenide heterostructures, Phys. Rev. B {\bf 96}, 041409(R) (2017).

\bibitem{SinEsp} S. Singh, C. Espejo, and A. H. Romero, Structural, electronic, vibrational, and elastic properties of graphene/WS$_2$ bilayer heterostructures, Phys. Rev. B {\bf 98}, 155309 (2018).

\bibitem{LopCol} A L\'{o}pez, L Colmen\'{a}rez, M Peralta, F Mireles, and E Medina, Proximity-induced spin-orbit effects in graphene on Au, Phys. Rev. B {\bf 99}, 085411 (2019).

\bibitem{PerMed} M. Peralta, E. Medina, and F. Mireles, Proximity-induced exchange and spin-orbit effects in graphene on Ni and Co, Phys. Rev. B {\bf 99}, 195452 (2019).

\bibitem{zhai2022} X. Zhai, Layered opposite Rashba spin-orbit coupling in bilayer graphene: Loss of spin chirality, symmetry breaking, and topological transition, Phys. Rev. B {\bf 105}, 205429 (2022).

\bibitem{LiZhang} L. Li, J. Zhang, G. Myeong, W. Shin, H. Lim, B. Kim, S. Kim, T. Jin, S. Cavill, B. S. Kim, C. Kim, J. Lischner, A. Ferreira, and S. Cho, Gate-Tunable Reversible Rashba-Edelstein Effect in a Few-Layer Graphene/2H-TaS$_2$ Heterostructure at Room Temperature, ACS Nano. {\bf 14}, 5251 (2020).

\bibitem{GhiKav2019} T. S. Ghiasi, A. A. Kaverzin, P. J. Blah, and B. J. van Wees, Charge-to-spin conversion by the Rashba-Edelstein effect in two-dimensional van der Waals heterostructures up to room temperature, Nano Lett. {\bf 19}, 5959 (2019).

\bibitem{BenTor} L. A. Ben\'{l}tez, W. S. Torres, J. F. Sierra, M. Timmermans, J. H. Garcia, S. Roche, M. V. Costache, and S. O. Valenzuela, Tunable room-temperature spin galvanic and spin Hall effects in van der Waals heterostructures, Nat. Mater. {\bf 19}, 170 (2020).

\bibitem{LuiLi} C. H. Lui, Z. Li, K. F. Mak, E. Cappelluti, and T. F. Heinz, Observation of an electrically tunable band gap in trilayer graphene, Nat. Phys. {\bf 7}, 944 (2011).

\bibitem{BaoJing} W. Bao, L. Jing, J. Velasco Jr, Y. Lee, G. Liu, D. Tran, B. Standley, M. Aykol, S. B. Cronin, D. Smirnov, M. Koshino, E. McCann, M. Bockrath, and C. N. Lau, Stacking-dependent band gap and quantum transport in trilayer graphene, Nat. Phys. {\bf 7}, 948 (2011).

\bibitem{JhaCra} S. H. Jhang, M. F. Craciun, S. Schmidmeier, S. Tokumitsu, S. Russo, M. Yamamoto, Y. Skourski, J. Wosnitza, S. Tarucha, J. Eroms, and C. Strunk, Stacking-order dependent transport properties of trilayer graphene, Phys. Rev. B {\bf 84}, 161408(R) (2011).

\bibitem{KhoKhr} T. Khodkov, I. Khrapach, M. F. Craciun, and S. Russo, Direct observation of a gate tunable band gap in electrical transport in ABC-trilayer graphene, Nano Lett. {\bf 15}, 4429 (2015).

\bibitem{ZolGmi2022} K. Zollner, M. Gmitra, and J. Fabian, Proximity spin-orbit and exchange coupling in ABA and ABC trilayer graphene van der Waals heterostructures, Phys. Rev. B {\bf 105}, 115126 (2022).

\bibitem{ShaLi} Y. Shan, Y. Li, D. Huang, Q. Tong, W. Yao, W. Liu, and S. Wu, Stacking symmetry governed second harmonic generation in graphene trilayers, Sci. Adv. {\bf 4}, 0074 (2018).

\bibitem{ZhangSahu} F. Zhang, B. Sahu, H. Min, and A. H. MacDonald,
    Band structure of ABC-stacked graphene trilayers, Phys. Rev. B {\bf 82}, 035409 (2010).

\bibitem{Kres} G. Kresse and J. Furthm\"{u}ller, Efficient iterative schemes for ab initio total-energy calculations using a plane-wave basis set. Phys. Rev. B {\bf 54}, 11169 (1996).

\bibitem{PerRu} J. P. Perdew, A. Ruzsinszky, G. I. Csonka, O. A. Vydrov, G. E. Scuseria, L. A. Constantin, X. Zhou, and K. Burke, Restoring the density-gradient expansion for exchange in solids and surfaces. Phys. Rev. Lett. {\bf 100}, 136406 (2008).

\bibitem{Ryb2018} A. G. Rybkin, A. A. Rybkina, M. M. Otrokov, O. Yu.
    Vilkov, I. I. Klimovskikh, A. E. Petukhov, M. V. Filianina, V. Yu. Voroshnin, I. P. Rusinov, A. Ernst, A. Arnau, E. V. Chulkov, and A. M. Shikin, Magneto-spin-orbit graphene: interplay between exchange and spin-orbit couplings, Nano Lett. {\bf 18}, 1564 (2018).

\bibitem{LeiTes2014} P. Leicht, J. Tesch, S. Bouvron, F. Blumenschein,
    P. Erler, L. Gragnaniello, and M. Fonin, Rashba splitting of graphene-covered Au(111) revealed by quasiparticle interference mapping, Phys. Rev. B {\bf 90}, 241406(R) (2014).

\bibitem{BaoYao} C. Bao, W. Yao, E. Wang, C. Chen, J. Avila, M. C.
    Asensio, and S. Zhou, Stacking-dependent electronic structure of trilayer graphene resolved by nanospot angle-resolved photoemission spectroscopy, Nano Lett. {\bf 17}, 1564 (2017).

\bibitem{LiZhang2022} Y. Li, S. Zhang, F. Chen, L. Wei, Z. Zhang, H.
    Xiao, H. Gao, M. Chen, S. Liang, D. Pei, L. Xu, K. Watanabe, T. Taniguchi, L. Yang, F. Miao, J. Liu, B. Cheng, M. Wang, Y. Chen, and Z. Liu, Observation of coexsiting dirac bands and Moir\'{e} flat bands in magic-angle twisted trilayer graphene, Adv. Mater. {\bf 34}, 2205996 (2022).

\bibitem{DevLed} T. Devakul, P. J. Ledwith, L.-Q. Xia, A. Uri, S. C.
    de la Barrera, P. Jarillo-Herrero, and L. Fu, Magic-angle helical trilayer graphene, Sci. Adv. {\bf 9}, eadi6063 (2023).

\bibitem{AviBand} Y. Avishai and Y. B. Band, Graphene bilayer and
    trilayer Moir\'{e} lattice with Rashba spin-orbit coupling, Phys. Rev. B {\bf 106}, L041406 (2022).

\end{thebibliography}
\end{document}